\documentclass[journal]{IEEEtran}

\ifCLASSINFOpdf
\else
\fi
\usepackage{graphicx}

\usepackage{balance}
\usepackage{setspace}
\usepackage{enumitem}

\begin{document}
%
\title{Can an AI-Powered Presentation Platform Based On The Game “Just a Minute” Be Used To Improve Students' Public Speaking Skills?}
%
%
%

\author{
    \IEEEauthorblockN{Frederic Higham\IEEEauthorrefmark{1} \quad Tommy Yuan\IEEEauthorrefmark{2}}

    \IEEEauthorblockA{\IEEEauthorrefmark{1} \IEEEauthorrefmark{2}
    Department of Computer Science, University of York, York, United Kingdom}

    \IEEEauthorblockA{Email: fh1061@york.ac.uk (F.H), tommy.yuan@york.ac.uk (T.Y)}

    \IEEEauthorblockA{\IEEEauthorrefmark{1} Corresponding author}
}

%
%

\markboth{}%
{}
%



\maketitle

\begin{abstract}

This study explores the effectiveness of applying AI and gamification into a presentation platform aimed at University students wanting to improve their public speaking skills in their native tongue. Specifically, a platform based on the radio show, Just a Minute (JAM), is explored. In this game, players are challenged to speak fluently on a topic for 60 seconds without repeating themselves, hesitating or deviating from the topic. JAM has proposed benefits such as allowing students to improve their spontaneous speaking skills and reduce their use of speech disfluencies (``um", `uh", etc.).

Previous research has highlighted the difficulties students face when speaking publicly, the main one being anxiety. AI Powered Presentation Platforms (AI-PPPs), where students can speak with an immersive AI audience and receive real-time feedback, have been explored as a method to improve student's speaking skills and confidence. So far they have shown promising results which this study aims to build upon. 

A group of students from the University of York are enlisted to evaluate the effectiveness of the JAM platform. They are asked to fill in a questionnaire, play through the game twice and then complete a final questionnaire to discuss their experiences playing the game. Various statistics are gathered during their gameplay such as the number of points they gained and the number of rules they broke. The results showed that students found the game promising and believed that their speaking skills could improve if they played the game for longer. More work will need to be carried out to prove the effectiveness of the game beyond the short term. 

\end{abstract}


\section{Introduction}

\IEEEPARstart{O}{ral} communication skills are vital throughout most aspects of our lives~\cite{PublicSpeakingToast}. In the case of university students, being able to speak effectively can help with class presentations, group projects, job interviews and networking. An important element of communication which many people struggle with is public speaking~\cite{PublicSpeakingAnxiety}. In this study, public speaking is defined as ``the structured way of an individual to speak directly to a group of people with an objective in mind of either informing, influencing, or entertaining them" \cite{PublicSpeakingDefinition}. This definition highlights the broad scope of public speaking, encompassing various communication contexts and purposes.

One of the main challenges students face when speaking publicly is anxiety, with one study showing that 75\% of Computer Science students face anxiety when speaking in public \cite{PublicSpeakingAnxiety}. Anxiety can take hold both before and during speech, leading to physical distress as well as a lower speech performance \cite{PublicSpeakingPrep}. This lower performance often manifests in rushed or fast spoken speech, possibly caused by a spike in adrenaline \cite{PublicSpeakingAnxiety}. It also leads to an increase in speech disfluencies such as filled pauses ("um", "uh", ...) \cite{hofmann1997speech}.

Artificial Intelligence Powered Presentation Platforms (AI-PPPs) have been explored as a promising tool to address the challenges students face in public speaking. By leveraging AI technologies, these platforms create immersive environments where users can practice speaking without the normal pressures of presenting in front of real humans. The users can also receive real-time, personalized feedback during their speeches to support their improvement.

One example of an AI-PPP involved using virtual reality (VR) to simulate a boardroom consisting of AI avatars. The avatars would react to the user's speech to provide feedback based on metrics such as the speaking speed and use of filler words by the user \cite{VirtualRealityExperiment}. This, as well as other studies, showed promising results for the effectiveness of the platforms' feedback. There were, however, concerns raised with how the feedback is produced and represented in AI-PPPs, particularly due to AI being inaccurate at judging human speech. In addition, few examples of these platforms have explored applying gamification (``the application of game elements in non-gaming contexts") to the user experience. 

In this study, an AI-PPP is developed which is inspired by the popular and long-standing BBC Radio 4 panel show game, ``Just a Minute" (JAM) \cite{wikipediaJustMinute}. In this game, four players compete for points by speaking on topics for as long as they can (up to 60 seconds) without repeating themselves, hesitating (pausing) or deviating from the topic. The aim is that, by developing a digital version of this game, the benefits of video games and gamification can be leveraged to provide students with an engaging platform where they can practice and improve their speaking skills. In addition, the format of JAM is expected to encourage students to avoid using speech disfluencies and improve their ability to speak spontaneously. 

The study investigates two central research questions (RQs), each containing sub questions:

\begin{enumerate}[label=RQ\arabic*), leftmargin=2.5em]
    \item Can an AI-PPP based on JAM be used to improve student's public speaking skills?
        \begin{enumerate}
            \item Does their anxiety become reduced?
            \item Do they become better at spontaneous speech? Do they learn to use less speech disfluencies? 
            \item Does the platform benefit from gamification. Do the students feel motivated to keep playing?
            \item Do the students feel immersed?
        \end{enumerate}
    \item  How effective is AI at powering this presentation platform?
        \begin{enumerate}
            \item Is the AI effective at evaluating, transcribing and producing spontaneous human speech for the purpose of this platform?
            \item How comfortable do the students feel when interacting with the AI in comparison to humans?
        \end{enumerate}
\end{enumerate}

These questions were constructed following the literature review and aim to cover only the most important aspects of this study needed to form a useful conclusion. Less of a focus is placed on speech anxiety as this has received the most attention in previous research on AI-PPPs and it is not a unique benefit provided by JAM. Instead, the most focus is placed on spontaneous speaking and speech disfluencies. 

Through answering these questions, the field of public speaking can be contributed to by demonstrating the effectiveness of applying gamification to AI-PPPs. Equally, a contribution is made to the study of JAM as a learning tool. 

\section{Literature Review}

This section aims to support the construction of the JAM AI-PPP by summarising the current state of public speaking research. Section \ref{good_public_speaker_section} discusses the public speaking skills that this platform should aim to improve and Section \ref{ppp_section} details the strengths and weaknesses found in previous AI-PPPs. The final section (\ref{JAM_section}), aims to link all of the previous research to JAM and justify the game's usage for this dissertation. 

\subsection{What Makes A Good Public Speaker? \label{good_public_speaker_section}}

Before developing a platform to improve public speaking skills, it is important to discuss what are considered good and bad practices when speaking. Various experts in public speaking have identified traits which make a good speaker. In general, the most important goals are to keep the audience engaged, speak clearly and appear credible \cite{PublicSpeakingCicero}. The following features and behaviours are discussed because of their importance to public speaking and their relevance to this study. Some features such a non-verbal language are not discussed because they are not directly relevant to the implemented platform. 

\subsubsection{Speech Disfluencies} \label{SpeechDsifluencies}

Speech disfluencies are features which cause a break or upset to fluent speech. This can include sentence changes, repetitions, stutter, omissions or sentence incompletion \cite{hofmann1997speech}. Filled pauses are considered one of the most significant disfluencies speakers make \cite{PublicSpeakingStrategies}. A filled pause (FP) is when the speaker substitutes a silent pause with a filler word (``um", ``uh", ``like") or with repeated words. This is thought to take place while the speaker is searching for a new word (lexical search phenomenon), particularly when the upcoming word or phrase is not frequent in the speaker's lexicon \cite{FillerWords}. It can also be caused by nervousness or divided attention. 

A study by Duvall et al. \cite{FillerWordsImpacts} found that the regular use of FPs can harm the credibility of a speaker. On the other hand, they found that using few FPs made a speech seem more scripted and, in some cases, less authentic. This is supported by the finding that scripted speeches tend to contain less FPs. A study by Strangert and Gustafson \cite{PublicSpeakingGoodSpeaker} found that speech disfluencies negatively correlated with scores allocated to speakers on their speaking ability. This could be due to a similar reason of people preferring non-scripted sounding speech. The reviewers were asked for opinions on disfluencies and they had mixed opinions. Around half of them viewed disfluencies as natural and unavoidable parts of speech whereas the other half viewed them as disruptive. 

FPs take place when there would otherwise be a silent pause. Giang~\cite{vinhgiangYearsCommunication} recommends using more silent pauses instead of FPs for two reasons. Firstly, it improves the clarity and credibility of the speech. Secondly, it gives the listeners more time to process what they have heard so far. A study from Zandan \cite{hbrStopSaying} shows that the average speaker uses 3.5 pauses per minute, with each pause consisting of a length of 0.2 -- 1.0 seconds. He states that the best speakers will pause more frequently than that and for longer (around 2 -- 3 seconds). 

It is clear that the presence of disfluencies (particularly FPs) within speeches is a nuanced topic. Public speaking can take place in many different contexts so the speaker must decide what approach is most appropriate. If a speaker were in a position where it was important to sound more authentic and honest, it could be within their interest to use some FPs. 

\subsubsection{Anxiety \& Confidence}\label{anxiety_section}

Anxiety is often cited as the main struggle with public speaking and studies have consistently found high rates of public speaking anxiety (PSA) amongst students. A study of 500 random residents in Winnipeg, showed that one third of people struggle with excessive anxiety when speaking to a large audience \cite{PublicSpeakingAnxietyStats}. This was explained by the following fears: ``Doing or saying something embarrassing (64\%), one's mind going blank (74\%), being unable to continue talking (63\%), saying foolish things or not making sense (59\%), and trembling, shaking, or showing other signs of anxiety (80\%)" \cite{PublicSpeakingAnxietyStats}. Another study of 50 Computer Science students found that 75\% of the students had PSA with 50\% of them saying it was due to a lack of confidence \cite{PublicSpeakingAnxiety}.

Anxiety can take hold both before and during a speech, leading to physical distress as well as a lower speech performance \cite{PublicSpeakingPrep}. This lower performance often manifests in a rushed or fast spoken speech, possibly caused by a spike in adrenaline \cite{PublicSpeakingAnxiety}. Hofman et al \cite{hofmann1997speech} found that speakers with PSA were more likely to use FPs and would use them for longer pauses. A few studies argue that some of the effects of anxiety on speaking can be seen as positive. For example, a study by Pope \cite{SpeakingInterviews} showed that anxiety could result in a more flustered speech but, at the same time, a higher productivity (more words spoken). 

The phobia of public speaking is an important problem to solve in the context of improving public speaking skills. Multiple solutions have been explored for it. It has been agreed upon that doing preparation, where students have a good idea of what they are going to speak about and can anticipate responses, is key to improving confidence \cite{PublicSpeakingLightning} \cite{PublicSpeakingAnxiety}. In addition, practising in less stressful environments, exposure therapy, is effective in gradually getting people used to public speaking \cite{PublicSpeakingExposureTherapy}. This could involve practising with smaller groups of people or using virtual reality (VR) technology. VR platforms have shown promising results for this problem \cite{PublicSpeakingExposureTherapy}. 



\subsubsection{Rhetoric Devices}

Rhetoric is the art of persuasion. Aristotle identified three core rhetoric appeals: logos, pathos, and ethos \cite{aristotle_rhetoric}. Logos involves the use of facts and figures to support the speaker's claims, pathos is the appeal to the listener's emotions and ethos is how the speaker demonstrates their authority and credibility. Many features of rhetoric such as quotations, repetitions, metaphors, personifications, analogies and inclusive language can be linked to these rhetoric appeals. Aristotle claimed that the best speakers would make use of all three of these appeals. Many public speaking professionals also place value in using rhetoric \cite{skillsyouneedUsingRhetorical}.

Liu et al \cite{TedTalkRhetoric} studied the links between the use of rhetoric devices and the applause received in TED Talks. Through analysing the data, they found that sentences with more personal pronouns were more likely to incite applause. This follows the idea that using inclusive language appeals to ethos by making the speaker feel more approachable. They also found that sentences which included ``numbers, certainty (e.g. always, never), and logical expressions" were conducive to applause, thus supporting the appeal of logos. The study is limited insofar that not every public speaking occasion involves or is searching for applause. That being said, it is still valuable in demonstrating the value of rhetoric in engaging with an audience.

\subsubsection{Review of Public Speaking Literature}

There are many resources available online which talk in detail about how to improve public speaking skills. While these resources are presented by people who speak publicly as a career, they often have little empirical basis and rely on anecdotal evidence. More research needs to be carried out to support the claims made by these professionals. 

There is also some disagreement on subjects such as disfluencies. This may be due to the subjectivity in what make a good public speech and the nuance in the different scenarios people speak in. Many of the advice given is more applicable to some scenarios than others. For example, the techniques you would want to deploy in a keynote presentation may differ to those in a job interview. It is important to consider the breadth of scenarios where public speaking takes place when doing research. 

\subsection{VR/AI Powered Presentation Platforms \label{ppp_section}}

Multiple researchers have developed and tested applications commonly referred to as VR/AI Powered Presentation Platforms (VR/AI-PPPs). These platforms aim to improve their users' public speaking skills and confidence through building an immersive interface for users to give speeches and receive real-time feedback. Palmas et al. \cite{VirtualRealityExperiment} created a VR application where users could give presentations in a simulated board meeting room. The audience consisted of interactive avatars whose behaviours would change based on the quality of the speech. The metrics they gathered to measure the quality of the speech included words per minute, eye contact, filler words and confidence amongst others. Their goal was to see if providing direct feedback during a user's speech would improve their experience compared to past applications. Half of the participants used a version with feedback while the other half received no feedback. The results showed that those who had feedback found the experience more motivating and were more comfortable using the technology. They did not choose to measure whether the application was effective in reducing anxiety. 

Cherner et al. \cite{VirtualRealityExperimentPitchVantage} experimented using a similar AI-PPP called Pitch Vantage \cite{pitchvantagePitchVantage}. Many of the participants felt that there was potential in being able to practice speaking but complained about the feedback they received after their speeches. They felt the feedback was inaccurate and not useful in improving their speeches. Using the app did not display any increase over time in their speaking skills. Beckner et al.~\cite{PitchVantageExperiment2} also carried out an experiment using Pitch Vantage but found more positive results. They tested the platform on civil engineering students and found that the students yielded significant improvements in their speaking ability. 

Tanveer et al. \cite{PublicSpeakingGoogleGlasses} carried out an experiment using the Google Glasses augmented reality display to give speakers live feedback during their speeches. In this case, the users were speaking in the real world, using the Google Glasses as a tool. The real-time feedback it provided included advice such as speaking slower and using less filled pauses. Their results showed that users found the feedback provided to be useful.

More research needs to be carried out to find if VR/AI Powered Presentation Platforms can reduce public speaking anxiety. The current research shows that the feedback provided by the platforms can be useful but that there is a risk of creating frustration in the users if the feedback is not implemented well enough. There has not been much exploration into gamified AI-PPPs. 

\subsection{Large Language Models (LLMs) \label{large_language_models}}

LLMs, are used to power many of the AI-PPP's features. In recent years, these models have progressed significantly in their ability to process and produce natural language \cite{ChatGPTJackofTrades}. OpenAI's ChatGPT \cite{openaiChatGPT} has most notably seen widespread use in a variety of applications. The model provides a powerful and easy to use API which can perform prompt based text generation, text-to-speech (TTS) and speech-to-text (STT) \cite{ChatGPTJackofTrades}. It allows you to produce a human-like speech on any chosen topic and then convert that speech to human-like audio. The user's speech can be understood, transcribed into text and then given as input to the text generation model to receive a response. It is worth noting that other models could have been used for the JAM platform, however, ChatGPT is one of the only ones to be powerful in both text and speech \cite{bumann2023chatbot}. It also has one of the most cost effective application programming interfaces. For this reason, it is the most convenient for development. 

Using AI in educational apps is becoming increasingly popular and successful. Khanmigo is a ChatGPT powered virtual tutor and coach, which helps students understand subjects by asking questions, giving hints, and guiding thought processes -- not just giving answers \cite{khanmigoAIpoweredTutor}. AI has seen use in many language learning apps such as Duolingo \cite{duolingo} where users can receive tailor fitted experiences. Studies have found these apps to be effective in their application of AI for personalised language learning. Despite the success ChatGPT has seen, there are currently some limitations when using it for this purpose:

\begin{itemize}
    \item ChatGPT is prompt-driven, which means any text generation requires a natural language prompt. Natural language is inherently imprecise, which sometimes makes getting a desired response difficult and unintuitive \cite{ChatGPTJackofTrades}. It can also sometimes be unpredictable in the format of its responses. 
    \item Applications which use prompt-driven models in the background are subject to prompt injection. This is where the front-end user is able to influence the prompts given to the model in the back-end (either intentionally or unintentionally), leading to unwanted consequences \cite{promptInjection}.
    \item ChatGPT can sometimes generate inaccurate responses with full confidence, which can mislead the user of the application \cite{ChatGPTJackofTrades}. This has been shown to cause frustration for users of AI-PPPs~\cite{VirtualRealityExperimentPitchVantage}. 
    \item Speech-To-Text is still not a polished technology. In the context of AI-PPPs, it has been known to provide inaccurate responses, particularly when the user's speech is not fully intelligible~\cite{VirtualRealityExperimentPitchVantage}. 
    \item Creating an application which is fully reliant on AI APIs can be expensive when used at scale and can be slow at producing responses (particularly for audio) \cite{predgenacceleratedinferencelarge}. This is especially pertinent in AI-PPPs which aim to simulate the dynamic nature of real life public speaking -- you don't want frequent pauses for loading.
\end{itemize}

These issues are factored into the development of the JAM platform. For example, care is taken to construct precise prompts so that the Chat GPT model produces consistent speeches for each of the AI speakers. 

\subsection{The ``Just a Minute" Game \label{JAM_section}}

\subsubsection{Rules and Format \label{JAM_rules_section}}

``Just a Minute" (JAM) is a BBC Radio Four panel comedy show which started in 1967 and has been running ever since \cite{wikipediaJustMinute}. Each episode features four panellists and one host. The players take it in turns to give a speech on a provided topic. Their goal is to speak for sixty seconds without breaking one of the following rules:

\begin{itemize}
    \item Hesitation: the speaker must not pause or use filler words for any longer than a brief moment in time. 
    \item Repetition: the speaker must not repeat a word more than once during their speech unless it is a common word (such as ``and") or the topic word. 
    \item Deviation: the speaker must not go off topic or speak in a way which does not make sense.
\end{itemize}

If a speaker breaks one of these rules, then one of the other players can raise a hand and state the rule which has been broken. The host makes a decision whether to accept this challenge or not. If the challenge is accepted, then the floor is given to the player who made the challenge. The player who is speaking when the sixty second timer ends is the winner of the round. The timer does not become reset when a new player gets control of the floor, meaning if a player makes a successful challenge thirty seconds into the first player's speech, then they only need to speak for thirty seconds more to win. 

The scoring works like so:

\begin{itemize}
    \item A player receives a point anytime they make a correct challenge. 
    \item If an incorrect challenge is made then the current speaker receives a point. 
    \item The speaker who wins the round gets a point.
    \item If the speaker who wins the round has spoken for the full sixty seconds (has not been challenged), they get an extra point.
\end{itemize}

The game is generally comedic in tone, with contestants giving funny speeches for each of the topics and occasionally making non-serious challenges as a joke. A more serious approach is taken in the implementation of this platform due to a more educational focus. 

\subsubsection{Relevant Literature}

Much research has been carried out on games (both physical and digital) for language learning and public speaking practice. However, JAM has received little individual focus, with only two relevant papers being published. These papers focus on using JAM as a classroom game for learning English as a second language. Jaelani and Utami \cite{JustaMinute1} carried out a survey amongst students who played JAM and found that the majority of students enjoyed the game, thought it was appropriate for the classroom and found it helped with their English speaking fluency. The second paper, by Rao \cite{JustaMinute2}, discusses proposed benefits such as improving student's ability to speak spontaneously without any fear and improving their listening skills. These papers are not completely relevant due to their focus on language learning instead of public speaking, however, they do demonstrate that JAM can be effective as a learning tool for students. They also showcase how the game can improve confidence in speaking which could translate to reducing speech anxiety. 


Jean-Pierre and Sturge \cite{PublicSpeakingLightning} experimented with using ``lightning talks" with school students. These talks were described as ``a short, time-limited, oral presentation on a subject that is completed without the use of supporting materials, such as PowerPoint slides, notes, or electronic devices". The students were given a few days to prepare and then they would give a talk in front of their class. Afterwards, the students were interviewed and gave feedback on their experience. The findings were as follows: ``Participants shared a belief that they developed their critical thinking skills, consolidated the flexibility and adaptability of their communication skills, enhanced their personal and professional confidence, and expanded their advocacy skills" \cite{PublicSpeakingLightning}. JAM could carry the same benefits of improving speaking confidence due to it following a similar format to these ``lightning talks". ``Just a Minute" was chosen for this study for multiple hypothesised reasons detailed in the next section.

\subsubsection{Public Speaking Benefits}

JAM encourages multiple good habits in public speaking. Firstly, it encourages speakers to avoid using filled pauses (FPs) due to the ``hesitation" rule. As was discussed in Section \ref{SpeechDsifluencies}, excessive use of FPs is a common speech disfluency and is seen as a bad practice. The players could become more self conscious of their use of FPs, the more they play the game due to them being challenged whenever a significant FP occurs. They could also pick up the habit of listening for FPs while other speakers have the floor. A drawback to the game is the fact that players are punished for using silent pauses as well as FPs. As was discussed in Section \ref{SpeechDsifluencies}, occasionally using silent pauses is encouraged in public speaking, especially as an alternative to FPs. To solve this problem, the rule of ``hesitation" is changed for this study. The player is allowed to make a few silent pauses per speech, in a similar way to how public speakers would pause for rhetorical effect.

The ``repetition" rule punishes players if they reuse any non-common word or topic word, making them think carefully about the vocabulary they use. As multiple instructors have illustrated, this is valuable in practising and improving language fluency. In terms of public speaking, the benefits are not so clear. There are some cases where repetitions can be used for rhetorical effect such as in anaphora where a word or phrase is repeated at the beginning of successive clauses. On the other hand, Jobe \cite{RepetitionRedundancy} discusses how repetitions in speech, when done badly, can often be redundant and ``get in the way without adding meaning". In addition, it is generally regarded as good practise to make use of synonyms and varied vocabulary when speaking or writing to make yourself sound more interesting \cite{justpublishingadviceSynonymsWriting}. Haines \cite{justpublishingadviceSynonymsWriting} discusses how the verb ``to get" can be used in place of many verbs and thus repeated often, however, that does not mean it should be used -- it would be better to use a variation of more specialised verbs. For example, instead of saying ``it's time to get more creative", you could say ``it's time to [become] / [start being] more creative". It is clear that repetition can be both detrimental and beneficial to public speaking depending on the context, whether or not the fact that JAM discourages repetition is beneficial to students needs to be explored. 


Finally, being encouraged to follow the rule of ``deviation" in JAM could be a good way to practice the ability to speak logically and to stay on topic instead of rambling. In addition, being concise and avoiding verbose speech is vital in many contexts -- listeners are more likely to lose attention if the speaker deviates. This is the most subjective of JAM rules so care needs to be taken when implementing it into a digital game. 

JAM challenges players to produce speeches on the spot with little or no preparation time. There are many cases in the real world where speakers must be prepared to speak on new and unfamiliar topics with short notice. For example, during a job interview, candidates need to be able to answer any question on any topic given to them with full confidence and no hesitation. The questionnaire's and interviews conducted as part of this study show that students find spontaneous speaking difficult and feel they have room for development. Not much research has been conducted on the subject of spontaneous speech so it is difficult to discuss its significance or how it can be improved. 


\subsubsection{Gamification and Digital Games}

JAM presents multiple benefits from gamification (``the application of game elements in non-gaming contexts" \cite{GameificationDefinition}). JAM can be seen as a public speaking exercise mixed with game elements. These game elements include, the interactivity from players making challenges during each other's speeches and the point gathering system. In this study, with JAM being in the form of a digital game (instead of physical), it is important to also discuss the application and benefits of digital games in learning. 

Using gamification design principles has been found to result in an increase in motivation for learning \cite{GameificationBenefits1}. One reason for this is the feedback digital games provide. In the case of JAM, each player is able to learn when they make a mistake and is rewarded for speaking well. This means JAM has the capacity to provide valuable feedback and motivation for its players. The VR-PPP study, by Palmas et al. \cite{VirtualRealityExperiment}, discussed in Section \ref{ppp_section}, found that implementing gamified and feedback elements was valuable in their application. 

Digital games benefit from learning through failure. Players can play and make mistakes without feeling peer pressure or embarrassment, which they otherwise could when playing with real players  \cite{lynch2017soft}. They also have the agency to choose when they want to play and for how long, which is especially valuable given the difficult requirements to set up a game of JAM (needs multiple players, a host, a place to play in, a timer etc.).

An important element of digital games is immersiveness. Brown and Cairns \cite{immersion} describes immersion as the experience of being drawn into a game, progressing through three levels: engagement (interest and attention), engrossment (emotional investment), and total immersion (where the game becomes the player's reality). Immersion is important in the context of a digital JAM game because the aim is to make the player feel like they are playing a real life match of JAM. Implementing human-like AI is expected to be an important factor in creating immersion. This is supported by the user interviews conducted for this study.

\subsection{Research Gaps}

This literature review identifies various gaps and room for development in the research of public speaking and AI-PPPs. These are summarised in the following questions:    

\begin{itemize}
    \item What characteristics make a good public speaker and what features should or should not be present in their speech? 
    \item How effective are AI-PPPs and VR applications at reducing a speaker's anxiety?
    \item How important are spontaneous speaking skills and do students struggle with speaking on the spot without a script or notes (like in JAM)?
    \item  Can JAM be used educationally, outside the context of learning English as a second language, such as in the context of improving public speaking skills?
\end{itemize}

Answering all of these questions is beyond the scope of the study. Instead a focus is placed on evaluating the effectiveness of AI-PPPs and gamification, through the medium of JAM. The research questions laid out in the introduction, are constructed with this in mind. 

\newpage
\section{Materials and Methods}

\subsection{The JAM Platform}

\subsubsection{Technologies Used}

The JAM platform is developed as a local-hosted web app. Flask is chosen as a framework due to its simplicity and integration with Python \cite{Flask}. The APIs for the ChatGPT models ``gpt-4o-mini" and ``tts-1" are used to power all of the AI functionality (speech generation, TTS and STT) for the reasons discussed in Section \ref{large_language_models}. The NLTK library in python is used for the text processing. No audio analysis takes place -- any speech from the user is converted to text via STT before being analysed. 

\subsubsection{Main Features Of The Platform}

\begin{figure}
    \centering
    \fbox{\includegraphics[width=8.5cm, height=4.5cm]{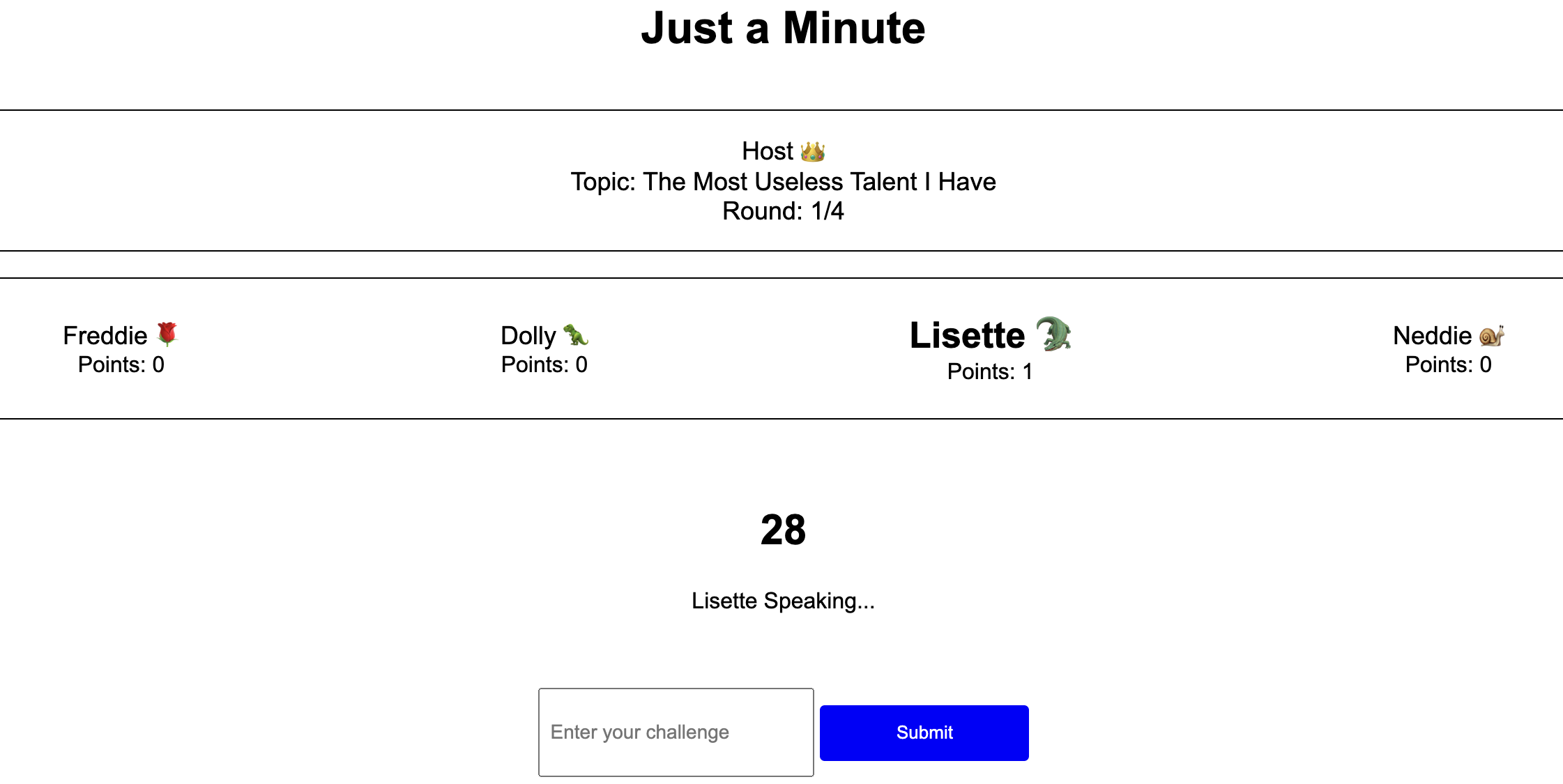}}
    \caption{A screenshot of the interface for the JAM game. This is the first round and Lissette is the current speaker. She has 28 seconds left on the clock.  \label{interface_diagram}}
\end{figure}

\begin{figure}
    \centering
    \fbox{\includegraphics[width=5.5cm, height=6cm]{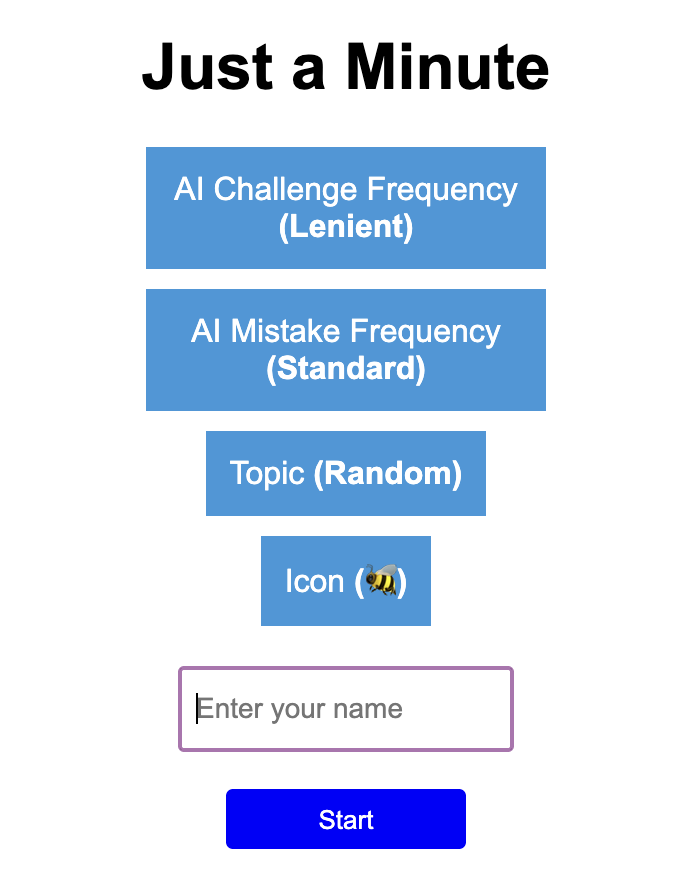}}
    \caption{A screenshot of the settings page for the JAM platform.  \label{settings_diagram}}
\end{figure}

 The following list describes the main features of the JAM platform. These were formed based on the user research and requirement gathering process. They aim to fully capitalise on the value of JAM and digital games in general.

\begin{itemize}
    \item Each round features randomly selected topics from a list. The user has the option to select their own list of topics in the settings or to use the pre-set topic list. Giving them the option to specify the topics allows them to tailor their experience to a particular area of interest or focus. For example, they may want to use the platform to practice answering interview questions.
    \item The user can select difficulty settings such as the frequency at which they will receive challenges. This allows the game to accommodate for a range of skill levels and avoids issues of frustration from the users (which have occurred in previous AI-PPPs). 
    \item Each AI speaker in the game has a randomized name, voice and personality. The personalities describe the kinds of mistakes the AI speakers are more likely to make as well as the frequency at which they will make challenges. These differences make the AI players feel more human-like and unique. An example of speech by the AI can be found in Appendix D.
    \item There is an AI host who introduces each round and announces the verdict on whether a challenge will be accepted or not. They also give a summary of the game at the end and congratulate the player with the most points. This makes the experience feel more lively and similar to the real game. 
\end{itemize}



\subsubsection{Generated Feedback}

At the end of each game, the user is given constructive criticism (generated by ChatGPT) for each of their speeches. This advice aims to help them to improve at playing JAM and giving entertaining speeches. It is one of the main ways that this platform aims to give feedback and improve its users' speaking skills over time. An example of this feedback can be found in Appendix C.

The feedback centres around tips on how to improve at JAM (such as how to avoid repetitions) and how to make the speech more entertaining (for example through use of rhetoric). It is important that this feedback rewards users for how engaging and entertaining their speech was because they are not necessarily rewarded for that during the game. 

\subsection{User Research \& Questionnaire} \label{user_research}

Before the JAM platform was developed, a user research and requirements gathering process was conducted to ensure the most appropriate features would be implemented. Five University of York students, who all demonstrated an interest in improving their public speaking skills, were interviewed in person. These students were asked various questions to gather details about what features the platform should include and to gather characteristics about the intended user-base. 

In addition to this, a questionnaire was distributed through various forums to a wider range of students from the University of York. It was filled out by 20 students and had the aim of learning about the students' experiences with public speaking and their opinions on the proposed JAM platform. As part of this, students were asked to watch a 2 minute gameplay video which showcases most of the platform's features and an AI player's speech. The questions from this questionnaire can be found in Appendix A. The results are discussed in Section \ref{Questionnaire_Results}. 

\subsection{Playtest Method}

Ten students from the University of York were enlisted to evaluate the effectiveness of the JAM platform. The process was as follows:

\begin{enumerate}
    \item Participants fill out the questionnaire discussed in Section \ref{Questionnaire_Results} detailing their current attitudes towards their public speaking ability and their expected performance at JAM.
    \item They play through the game twice in close succession which, on average, amounts to 20 minutes of gameplay. Each participant plays through the same set of topics as each other. The following statistics are gathered for each round and game:
    \begin{itemize}
        \item The number of hesitations, repetitions and deviations they made.
        \item The number of challenges they made and received.
        \item The points they received.
        \item Whether they won or lost a round, and by how many points.
    \end{itemize}
    \item They are asked to look at the transcriptions for each of their speeches and identify the number of incorrectly transcribed words or punctuation there are. 
    \item They fill out a post-playtest questionnaire which requires them to reflect on their experience playing the game. These questions can be found in Appendix B.
    \item They are asked verbal questions which invite them to expand upon their answers in the survey. 
\end{enumerate}

By having a survey before and after, it is possible to compare and contrast their feelings after playing the game. Most of the survey questions invite participants to give a rating from 1-10 on their feelings regarding public speaking and the JAM game. This is a simple method which gives them an adequate amount of flexibility to express their feelings \cite{surveys}. It also allows for quantitative analysis of the data which is easier to summarise and describe amongst all the participants compared to qualitative data. The verbal discussion after the surveys allows the participants to provide any qualitative information which they weren't able to express through the surveys. 

Gameplay statistics are gathered in order to gain more insight into the abilities of the students and their experience playing the game. For example, if students repeatedly break the rule of hesitation due to pausing, this can be seen as evidence for students struggling with spontaneous speech. Due to the evaluations taking place over short periods, it is not possible to analyse any long term improvements in performance for the students. The aim is that, in the future, a longer term study can be conducted and the performance statistics can be leveraged to demonstrate an improvement in the students' speaking skills. The following formulas could be used as part of this.

A \emph{performance score} can be calculated for each round using the formula:
\[PerfomanceScore = UserPoints - AIPoints\]
This formula represents a ratio for the number of points the player has gained (this round only) against the number of points other players (the AI) have received. The goal is to finish with a higher number of points than the opponents, therefore this is a useful representation of performance for any given round. In addition a \emph{rules broken} percentage can be calculated for each speech using the formula:
\[RulesBroken = \frac{(Hesitations + Repetitions + Deviations)}{SpeechLength}\]
This represents the percentage of a speech containing rule breaking and provides a way to analyse the performance quality of a speech -- the best speeches in JAM will minimise this percentage. \emph{Hesitations} is counted as the number of distinct hesitation units in a speech, \emph{Repetitions} is counted as the number of unique words repeated more than once and \emph{Deviations} is counted as the number of incorrect topics discussed. The \emph{SpeechLength} is the number of word tokens in a speech. 

\section{Results and Discussion}

\subsection{Questionnaire Results} \label{Questionnaire_Results}

The questionnaire is split into two sections: the ``public speaking background" and then the ``JAM questions". The results for the first section are summarised as follows:

\begin{itemize}
    \item Students, on average, gave themselves a score of 4/10 for their public speaking ability and 7/10 for the amount of anxiety they experience.
    \item 90\% of students described public speaking as being an important subject to them and something they wanted to improve upon.
    \item Students gave themselves a score of 6.5/10 on average for their ability to speak spontaneously and answer questions on the spot. In addition, they scored themselves at 7.5/10 for the frequency of filled pauses they used.
\end{itemize}

These results indicate that students are in a position to be able to benefit from any AI-PPP, such as this one, which aims to improve spontaneous speaking and reduce speech disfluencies. It also indicates that they have an incentive to use any such platform, given the improvement of their speaking skills is important to them. Previous studies about public speaking and speech anxiety have found similar results.

The results for the JAM questions are as follows:

\begin{itemize}
    \item Most students have not heard of JAM before but are interested in playing it. Those who have listened to the show or played the game, had an enjoyable experience. For those who were not familiar with the game, 75\% of them demonstrated interest in playing it. 
    \item 60\% of the students did not believe they could speak for the full 60 seconds in JAM without breaking a rule. 
    \item 95\% of the students believed playing JAM in-person with humans could allow them to improve their speaking skills.
    \item Only 45\% of students believed that using this platform could reduce their anxiety, however, 90\% believed it could improve their spontaneous speaking skills and reduce their use of filled pauses. 
    \item 65\% of students responded with ``somewhat" when asked if the AI speakers sounded realistic and 20\% responded with ``yes". 
\end{itemize}

These results further indicate an interest amongst the students in using this AI-PPP and playing the game of JAM. The students also believe that this platform could be beneficial for their spontaneous speech and use of filled pauses. However, not many students believe their anxiety could be improved. This could be to do with the abstract and non-human nature of this platform -- they don't get enough of the opportunity to engage in exposure therapy because they don't feel exposed to the full experience of public speaking. Previous studies into AI-PPPs involved avatars and 3D graphics which could explain their better results on reducing anxiety. The majority of students believed that the AI sounded at least ``somewhat" realistic in terms of their human-like voices and the content of their speech. This demonstrates the power of ChatGPT's TTS capability as well as its ability to produce speeches for JAM. 

It is important to note that the opinions about the platform from the respondents are speculative as they did not have the opportunity to play the game, however, they were shown a gameplay video. 

\subsection{Post-Playtest Questionnaire \& Discussion Results}

In the post-playtest survey, 90\% of students responded saying that they enjoyed playing the game and that they would play it again. This supports the research question (RQ) 1c (the benefits of gamification on motivation and enjoyment) because it shows they are motivated to keep playing. Participants scored the immersiveness of the game at 6.5/10. From the discussions held after the survey and the questionnaire discussed above, it is clear that there is some room for improvement in the immersiveness. Some students suggested that introducing avatars for the AI players could help as well as reducing some of the loading times. Nevertheless, the results suggest that RQ 1d has been supported. 

Seventy percent of participants found the feedback from the game at least somewhat helpful. This supports RQ 2a (the effectiveness of the AI evaluations) because it indicates that the AI is intelligent enough to provide clear and constructive criticism. Sixty percent of participants responded with at least "somewhat" when asked if the challenges made to them by their opponents were fair. A slightly higher percentage of students responded with ``yes" when asked if their own challenges were judged fairly. Sometimes the students would be challenged for hesitating or repeating when they didn't feel like they deserved it. For example, one student was challenged for repeating the word ``like" and felt like the word was too common to be counted as a repetition word. There is some subjectivity when it comes to deciding JAM challenges and some of the students felt frustrated when the AI made a decision they disagreed with. It follows that RQ 2a and 2b are only partially supported. 

Every person who played the game felt more confident afterwards in their ability to play JAM. This is particularly notable given a large percentage of them showed concerns with how well they would do to begin with. When asked if they felt like their ability at public speaking and speaking spontaneously has increased, most responded with ``maybe". The same was found for anxiety. This makes sense as it would be hard to judge if your speaking skills had improved after only 20 minutes of gameplay. It could also suggest that the platform is not effective at improving skills and anxiety, however the rest of the data suggests otherwise. For example, 80\% responded with ``yes" when asked if playing the game more times in the future could improve their speaking skills. The same was found for anxiety. This is promising and suggests that RQ 1a and 1b could be supported if the game were played on a longer-term basis. 

\subsection{Gameplay Statistics}

The transcription of the user's speeches was not always accurate. The statistics gathered showed that 35\% of word tokens were incorrectly recognised. Sometimes it would replace a word with one of its homophones or just completely misunderstand the word. It would also use punctuation in an unnatural way such as placing full stops in the middle of a sentence. The correct use of punctuation is important as this is factored into the analysis of the speeches. Another issue is that sometimes ChatGPT hallucinates whenever it receives a period of silence to transcribe -- it will produce phrases which were never said such as ``thanks for watching" or ``goodbye". This is a known problem and is likely due to the model being trained on Youtube videos where these phrases are often spoken \cite{openaiMisinterpretationNonspeech}. It is possible that on some occasions, the students were challenged during their speech as a direct result of an incorrect transcription. When asked on their opinions regarding these errors, most participants felt that the occasional incorrect word did not harm their experience. A possible solution was discussed which involves allowing the user to appeal incorrect challenges or to manually modify the transcription of their speech. It is worth noting that ChatGPT is continuously improving its STT capabilities and that many of these issues may be solved. 

On average, students broke the rule of hesitation at least once in their speech. The most common cause was due to a silent pause as opposed to the use of filled pauses (FP). FPs could still be found in most speeches, however, they were infrequent (around one per speech) and normally did not pass the threshold of being a hesitation. It seems students believe that they use FPs more often than they actually do, looking at their responses to the questionnaire compared to their actual speeches. It is worth noting that they likely used less FPs whilst playing the game than they normally would, given the rules of the game. This could suggest that the platform is working as intended -- that the students are successfully able to reduce their use of FPs by playing the game. Whether this can impact their use of FPs in other public speaking contexts is yet to be shown. 

Students also broke the repetition rule as often as they did the hesitation rule (once per speech). This is a hard rule to follow so it is expected that it will be broken at least once. The deviation rule was never broken (except by one student who broke it on purpose) which makes sense given it is a rule which mainly exists to ensure the topic is kept to as opposed to catching players out. 

\subsection{Summary}

\begin{enumerate}
    \item \textbf{RQ 1a-b:} This study does not provide quantitative proof that using the platform can improve the students' speaking skills. However, the questionnaires filled out by the students show promising results. The majority of students who used the platform found it enjoyable and believed that playing it more often could improve their speaking skills. The participants who only watched a gameplay video also believed that the platform could improve their skills. A longer term study using the gameplay statistics could be conducted to help prove this RQ. 
    \item \textbf{RQ 1c-d:} The platform benefiting from gamification has been supported. Most participants found the experience fun and engaging. They did not show frustration when failing (except when they received challenges that they thought were unfair). They found the evaluations of their speeches to be helpful, aiding the process of ``learning through failure". Most students found the game at least somewhat immersive. The interface could be less abstract and feature more realistic graphics to further benefit from gamification.
    \item \textbf{RQ 2:} The chosen AI was effective at generating human-like speeches and producing human-like voices but was not always effective at transcribing the users' speeches. That being said, the students did not feel like this impacted their experience. The constructive criticisms that the AI produced for the users' speeches were insightful and helpful to them.
\end{enumerate}

\section{Conclusion}

This study set out to evaluate the potential of an AI-PPP modelled on ``Just a Minute" to improve students’ public speaking skills. The research focused specifically on whether such a platform could enhance spontaneous speaking, reduce speech disfluencies, and provide an engaging and immersive user experience through gamification and AI feedback.

The results of the short-term user study demonstrated that students responded positively to the game. Most participants found the experience enjoyable and expressed a willingness to continue playing, suggesting that the gamified format can significantly boost motivation. While no measurable improvement in speaking skills or anxiety levels was established over the brief duration of the trial, the majority of participants believed that repeated use of the platform could lead to such improvements over time. The AI-generated feedback was generally perceived as helpful, though transcription inaccuracies occasionally impacted the fairness of gameplay challenges.

The study contributes to the emerging field of AI-PPPs by demonstrating how structured game-based exercises, especially those focusing on real-time rules like JAM, can support the practice of key public speaking skills in a low-pressure, repeatable environment. It also highlights current limitations of AI technologies such as STT accuracy and latency, especially in audio contexts, which can affect both performance and immersion.

Future research should focus on longitudinal studies with more participants to measure sustained improvement in public speaking ability and reductions in anxiety. Enhancements such as avatar-based AI speakers and improved real-time feedback mechanisms may also increase immersion and perceived realism. Despite its limitations, this study provides encouraging evidence that AI-PPPs enhanced by gamification can play a meaningful role in public speaking education. 




\section*{Appendix}

\subsection{User Research Questionnaire} \label{questionaire}

Public speaking questions:
\begin{enumerate} 
    \item How good do you feel you are at public speaking? (1-10)
    \item How much anxiety do you feel when taking part in public speaking? (1-10)
    \item Do you feel a desire to improve your public speaking skills? (yes/no/maybe)
    \item How important are public speaking skills to you? (1-10)
    \item How confident are you when answering questions on the spot (such as during an interview)? (1-10)
    \item How often do you use filled pauses (``uh", ``um", ...) when you are engaging in public speaking? (1-10)
    \item If you had to give a short impromptu speech without any notes in front of a small audience, how well do you think you would cope? (yes/no/maybe)
\end{enumerate}

JAM questions:
\begin{enumerate}
    \item Have you watched or listened to the show ``Just a Minute" before? (yes/no/maybe)
    \item Have you played ``Just a Minute" before? (yes/no/something similar)
    \item Do you think you would enjoy playing ``Just a Minute"? If you have played it, did you enjoy doing so? (yes/no/maybe)
    \item Do you think you can speak for 60 seconds in the game of ``Just a Minute" without repeating, deviating or hesitating? (yes/no/maybe)
    \item Do you think playing ``Just a Minute" in-person with a group of people could improve your public speaking skills? (yes/no/maybe)
    \item Do you think an app based on ``Just a Minute" where you compete with AI players would be interesting to you? (yes/no/maybe)
    \item Do you think playing this JAM app could reduce your public speaking anxiety? (yes/no/maybe)
    \item Do you think playing this JAM app could reduce your use of filled pauses (``um", ``uh", ...)? (yes/no/maybe)
    \item Do you think playing this JAM app could improve your ability to speak spontaneously / on the spot? (yes/no/maybe)
    \item Looking at the gameplay video provided, do you think the AI speakers seem realistic in terms of what they are saying and how their voices sound? (yes/no/somewhat)
\end{enumerate}

\subsection{Post-playtest Questionnaire} \label{questionaire}

\begin{enumerate}
    \item How much did you enjoy playing the game? (1-10)
    \item Would you play the game again? (Yes/No/Maybe)
    \item How immersed were you in the game? (1-10)
    \item Did the game provide useful feedback? (Yes/No/Somewhat)
    \item Did you feel like the challenges made to you by your opponents were fair? (Yes/No/Somewhat)
    \item Did you feel like the challenges you made were judged fairly? (Yes/No/Somewhat)
    \item Do you feel more confident at playing Just a Minute now? (Yes/No/Maybe)
    \item Do you feel less anxiety about public speaking after playing Just a Minute? (Yes/No/Maybe)
    \item Do you think playing this game has made you more capable at public speaking and speaking spontaneously? (Yes/No/Maybe)
    \item Do you think playing this game \textbf{more} times could make you more confident at public speaking and speaking spontaneously? (Yes/No/Maybe)
    \item Do you think playing this game \textbf{more} times could make you less anxious about public speaking? (Yes/No/Maybe)
\end{enumerate}

\subsection{Example Feedback}

This is an example of feedback provided to a student about their speech on the topic of ``A Day In The Life Of My Pet":

\begin{quote}
   ``This speech has the potential to be delightful, especially for pet lovers. However, it lacks depth and variety. Instead of repeating phrases like `he likes to eat and he likes to sleep', consider incorporating anecdotes or humorous stories about your dog, which could elicit laughs or smiles from the audience. You might say, `Shaun loves to eat, but have you seen the way he devours a piece of bacon? It's like watching a tornado in action!' Such descriptions can add personality and make the speech more relatable. Also, remember to include sensory details. Mention textures, sounds, or specific places during hikes that could paint a clearer picture for the audience." 
\end{quote}

\subsection{Example Speech}

The following is an example of a speech generated for one of the AI speakers in the game. It is on the topic of ``The Most Embarrassing Moment Of My Life" and the speaker was set to produce frequent rules breaking (shown in bold).

\begin{quote}
``So, uh, the most embarrassing moment of my life\textbf{… hmm, }it’s tough to pick just one, but there was this time in high \textbf{school}, right? I was in the cafeteria\textbf{, and I, um,} thought it would be funny to try and impress my friends by, you know, balancing a \textbf{milk} \textbf{carton} on my head. I mean, who hasn’t tried that at some point? But as I was standing there, feeling all cool and casual, well, I lost my balance and\textbf{...} boom! It fell and, like, exploded all over me—white \textbf{milk} everywhere\textbf{! And, um, }the worst part? Everybody just burst out laughing,\textbf{ and I was, like, }totally mortified. My best friend Sarah couldn’t stop giggling and even started making jokes about \textbf{how I should, uh, join a circus instead of being at }school\textbf{, which is totally off-topic, I know, but you get what I mean, right?} Just, like, picture me covered in \textbf{milk}, looking like a complete disaster. Oh, and then I had to walk home, you know, feeling sticky and kind of\textbf{...} ugh, gross. I think that was the day I learned a valuable lesson about just sticking to the sidelines instead of trying to be a, uh, clown in front of everyone\textbf{. So yeah,} that was definitely my most embarrassing moment—let’s just say I never tried to impress anyone with a \textbf{milk} \textbf{carton} again!"
\end{quote}

\section*{Conflict of Interest}

The authors declare no conflict of interest.

\section*{Author Contributions}

Freddie Higham wrote the paper and conducted the study. Tommy Yuan supervised the project, proofread the paper and gave advice. 


%


\ifCLASSOPTIONcaptionsoff
  \newpage
\fi



%

\bibliographystyle{IEEEtran}
\bibliography{bibtex}

%





\end{document}